\documentclass[12pt]{article}
\title{\bf Resonance radiative decays \\ as a tool for its parity
determination.}
\author{{B.L.\,Ioffe and A.V.\,Samsonov}\\
\\ {\small\it Institute of Theoretical and Experimental Physics,}\\
{\small\it B.Cheremushkinskaya, 25, 117218, Moscow, Russia}}
\begin{document}
\date{}
\maketitle
\newcommand{\be}{\begin{equation}}
\newcommand{\ee}{\end{equation}}
\def\la{\mathrel{\mathpalette\fun <}}
\def\ga{\mathrel{\mathpalette\fun >}}
\def\fun#1#2{\lower3.6pt\vbox{\baselineskip0pt\lineskip.9pt
\ialign{$\mathsurround=0pt#1\hfil##\hfil$\crcr#2\crcr\sim\crcr}}}

\vspace{1cm}
\begin{abstract}
Radiative decays of the spin 1/2 baryonic resonances $R$ with the 
decay mode $R\to KN$ in case of  small energy release are considered.
Pentaquark $\Theta^+$ is an example of such resonance. 
It is shown that in case of
positive resonance parity $(J^p=1/2^+)$ 
corrections to the soft photon radiation formula are large even at low
photon energies $\omega \ga 20$ MeV and structure terms
contributions may be  essential, if $R$ size $r_0 > 1$ fm.
This effect is absent in case of negative  
parity $(J^p=1/2^-)$. Particularly, measurements of
the $\gamma$-spectrum in $\Theta^+$ radiative decays may allow us to
determine $\Theta^+$ parity.
\end{abstract}

\newpage
\indent In the present paper we show that the parity of the spin 1/2 
baryonic resonance $R$ with the 
decay mode $R\to KN$ (kaon and nucleon)
in case of the small energe release can be determined 
by study its radiative decays $R\to KN\gamma$. A widely known example of
such resonance is pentaquark $\Theta^+$. So, for definiteness we consider 
$\Theta^+$ decays, nevertheless, our results are general and refer to any 
resonance with mentioned above properties. \\
\indent An exotic baryonic resonance $\Theta^+$ with the mass
$m_{\Theta}=1540$ MeV and quark content $uudd\bar{s}$ had been
found two years ago [1,2]. Later, the existence of $\Theta^+$
was confirmed in many experiments. The main decay modes of
$\Theta^+$ are $\Theta^+\to pK^0$ and $\Theta^+\to nK^+$.
Experimentally, the upper limit on $\Theta^+$ width was found
$\Gamma > 9$ MeV [2]. From the phase analysis of KN scattering
[3] and theoretical analysis of $\Theta^+$ production
mechanism in $K^+d\to p\,pK^0$ [4] and $K^0Xe$ [5]
reactions more strict limitations on $\Gamma$ were found: 
\\ $\Gamma\leq 1$ MeV. The parity of $\Theta^+$ is unknown  experimentally.
From theoretical point of view, the positive parity is favoured,
since at $P=+1$ and supposed spin $J=1/2$, $KN$ are produced in
$P$-wave in $\Theta^+$ decay, and  the
narrow $\Theta^+$ width can be understand easier. 
Since the decays $\Theta^{++}\to pK^+$ were not found, the $\Theta$ isospin 
is zero. \\
\indent $\Theta^+$ baryon was predicted by Diakonov, Petrov and Polyakov
[6] in Chiral Quark Soliton Model (CQSM) as a member of
anti-decuplet -- a rotational excitation in colour-flavour space
in CQSM. In this case the $\Theta^+$ parity is +1. However, CQSM
does not explain the unusual very narrow $\Theta^+$ width. The
explanation of $\Theta^+$ narrow width based on chiral
conservation [7] (where it is supposed that $\Theta^+$ is a
compact object), is valid for any $\Theta^+$ parity $P=\pm 1$.
Other models also allow $P=-1$ for $\Theta^+$. Therefore, the
measurement of $\Theta^+$ parity is desirable. \\
\indent We consider the radiative decays of pentaquark
$\Theta^+\to p\,K^0\gamma$, $\Theta^+\to n\,K^+\gamma$ and show that the
$\gamma$-spectra in radiative decays are essentially different
in cases of positive and negative pentaquark parities, what would 
allow us to determine $\Theta^+$ parity, basing on radiative decay data. \\
\indent Consider the emission of soft gamma's with energies  $\omega \leq
50$ MeV. Since the total energy  release in $\Theta^+\to KN$ decay
is 100 MeV, the main part of $\gamma$ spectrum is just in
this domain. The wave lengths  of gamma's in this domain are
larger than 4 fm, and one may expect that they are larger than
$\Theta^+$ size. So, the general formula for accompanying photon
emission in decay process can be used:
\be dW_{\gamma}(\omega) = \frac{2\alpha}{3\pi}
\frac{p^2}{E^2_{ch}}\frac{d\omega}{\omega}
W_{pK^0,nK^+}.\label{1}\ee 
Here $\alpha=1/137$, $p$ is the $N$ or $K$ momentum 
in the $\Theta^+$ rest system, 
$E_{ch}$ is the total energy of the charged
particle in the final state (i.e. proton or $K^+$ meson) and
$W_{pK^0}$, $W_{nK^+}$ are the probabilities of $\Theta^+\to pK^0$
and $\Theta^+\to nK^+$ decays (which are approximately equal). In
(1) it was put approximately $E_{ch}\gg p$. This formula
is a general relation, corresponding to the case when charged
particle starts to move suddenly and for this reason emits
photons. Equation (1) can be derived classically. (Originally,
it was obtained by Pomeranchuk  and Shmushkevich for  
photon emission in charge exchange $n-p$ scattering [8]. In
[9] formula (1) was derived for $\pi\to \mu\nu\gamma$ decay
and the general character of this equation was mentioned). \\
\indent In case of $\theta^+$ positive parity the final particles in
$\Theta^+\to NK$ decay are produced in the state with orbital
momentum $L=1$, what results in strong suppression of the decay rate.
The photon has spin 1 and negative parity (in case of electric field). 
Photon emission
takes off this suppression, what leads to relative enhancement of
$\gamma$-radiation. For this reason one may expect
large corrections to (1) even at low $\omega$. No such
effect exists for negative $\Theta^+$ parity, where only small
corrections to (1) take place. \\
\indent Let us calculate this effect quantitatively. The
phenomenological $\Theta^+$ decay Hamiltonian is supposed to be
\be H_{int}
=f\bar{\psi}_N(i\gamma_5,1)\psi_{\Theta}\varphi_K+c.c.,\label{2}\ee
where $i\gamma_5$ and 1 in the brackets correspond to the positive and
negative $\Theta^+$ parities, $\psi_N$ and $\varphi_K$ are
isospinors and their product is an isoscalar. From (2) we
get for the widths in cases of positive and negative $\Theta^+$
parities:
\be
P=+1:
~~~~~~~~~~~~~\Gamma=2f^2(E_N-m_N)\frac{p}{m_{\Theta}},~~~~~~~
f^2=0.083\,,\label{3}\ee
\be
~~~~~P=-1:~~~~~~~~~~~~~\Gamma=4f^2
m_N\frac{p}{m_{\Theta}}\,,~~~~~~~~~~~~~~~~~ f^2=1.6\cdot 10^{-3}.
\label{4}\ee 
In (3), (4) $E_N$ and $m_N$ are the energy and mass 
of the nucleon. The values of the effective
coupling constants  $f^2$, corresponding to $\Gamma=1$
MeV, are also shown. \\
\indent 
In calculation of the radiative $\Theta^+$ decay, besides the standard
Feynman diagrams, describing the photon emission by initial and
final charged particles, we account the structure dependent term, where 
photon is emitted during the decay process. The effective Hamiltonian of
this term is assumed to be
\be
H_{str}=ge\bar{\psi}_N\gamma_{\mu}\gamma_5\psi_{\Theta}\frac{\partial
\varphi_K}{\partial x_{\nu}}F_{\mu\nu} + c.c.\,,\label{5}\ee 
where $F_{\mu\nu}$ is the electromagnetic field strength, and
$g=g_{pK^0}$ for $\Theta^+\to pK^0\gamma$, $g=g_{nK^+}$ for
 $\Theta^+\to nK^+\gamma$. For the quantitative
estimations we consider only this simplest form of $H_{str}$. 
The differential
probabilities of radiative decays $\Theta^+\to pK^0\gamma$ and
$\Theta^+\to nK^+\gamma$ in case of positive $\Theta^+$ parity
were found to be:
\be dW_{pK^0\gamma} = \frac{2\alpha}{3\pi}\,\frac{p^2}{E_p^2}\,
\frac{d\omega}{\omega}B_{pK^0}(\omega) W_{pK^0}\,,\label{6} \ee
$$\displaylines{B_{pK^0}(\omega)=1+\frac{\omega}{E_p-m_p} 
\biggl(1-\frac{m_p}{m_\Theta}\biggr) + 
\frac{3}{2}\,\frac{m_p\omega^2}{p^2(E_p - m_p)} 
\biggl(1-\frac{m_p}{m_\Theta}\biggr)^2-\hfill}$$ 
$$\displaylines{\hfill-2\,\frac{g_{pK^0}}{f}
\frac{m_\Theta-m_p}{E_p - m_p}\,m_p\,\omega^2\,,~~~~(7)}$$
$$ dW_{nK^+\gamma} = \frac{2\alpha}{3 \pi}\frac{p^2}{E_K^2}~
\frac{d \omega}{\omega}~ B^{(\omega)}_{nK^+} W_{nK+}\,, \eqno(8)$$
$$ B_{nK^+}(\omega) = 1 + \frac{\omega}{E_n-m_n}\,\frac{m_K}{m_\Theta}+
\frac{3}{2}\,\frac{m_n \omega^2}{p^2(E_n - m_n)}\frac{m_K^2}{m_\Theta^2}
+\,2\,\frac{g_{nK^+}}{f}
\frac{m_\Theta-m_n}{E_n-m_n}\,m_K\,\omega^2\,. \eqno(9)$$
In (6)-(9) $p$ was neglected in comparison with $E_p$,
$E_k$ and the decrease of $(N,K)$ phase space because of the
photon emission was disregarded. Only the terms linear in $g$ are
retained. The photon emission due to magnetic moments of $\Theta^+$ and 
nucleon is not enhanced and small. Indeed, the magnetic interaction has 
another $P$-parity in comparison with the electric one, therefore, the 
suppression remains. In the corresponding terms in (7), (9)
factor $E_N-m_N$ in the denominator is absent. For this reason we omit 
these terms. \\
\indent Numerically, in $\Theta^+ \to KN$ decay $p = 260$\,MeV,
$E_p - m_p \approx E_n - m_n = 36$\,MeV. At $\omega = 50$~ MeV the
correction terms (without the structure terms) are equal to 0.76
in $B_{pK^0}$ and 0.60 in $B_{nK^+}$. Therefore, the deviation
from the soft photon radiation formula (1) is noticeable in
these decays. \\
\indent The structure constants $g$ 
may be estimated as
$$ g \sim 2\,\frac{1}{m_\Theta+m_d}\,\frac{1}{m_\Theta}\,r_0\,, $$
where $r_0$ is the $\Theta^+$ size and factor
2 follows from comparison with the $\bar{N}NK$  coupling constant
$g^2_{\bar{N}NK} \sim 5$. If $r_0 \sim 1$ fm, the last terms
in (7), (9) are of order 0.3-0.5 at $\omega = 50$\,MeV.
The observation of this term  allows one to estimate pentaquark size. \\
\indent The conclusion is, the photon spectra in radiative $\Theta^+$
decay are essentially different from one another in cases of
positive and negative $\Theta^+$ parities. For negative $\Theta^+$
parity the spectrum in the domain of low energy photons is well
described by the soft photon emission formula (\ref{1}), while for 
positive $\Theta^+$ parity the deviations from (\ref{1}) are
noticeable. Unfortunately, the radiative decay branching
ratio is low: for photons in the energy interval $\omega =
10-50$~MeV it is about $(0.3-1.0)\,10^{-3}$. \\
\indent The $\Theta^+$ radiative decays were considered in [10, 11]. 
In [10] the 
strong mixing of anti-decuplet with octet was assumed and very large 
branching ratio $\sim 3\cdot10^{-2}$ was found. However, the results are
not reliable, since during the calculation the gauge invariance of 
electromagnetic interaction was violated. [11] has some resemblance to our
paper, but photon spectrum was not analysed and the main issue of
our paper -- the difference of spectra in cases of positive
and negative parities -- was not found. Moreover, in [11] it was claimed that
the radiative decay rates are equal for both $\Theta^+$ parities. This
statement arises, because the authors choose very small low limit of 
integration over photon energy, which could not be achieved experimentally.
\\\\
\indent We are thankful to Yu.S.\,Kalashnikova for useful remarks. \\
\indent This work is supported in part by US Civilian Research and
Development Foundation (CRDF) Cooperative Grant Program, Project
RUP2-2621-MO-04, RFBR grant 03-02-16209 and by the funds from EC
to the project "Study of Strongly Interacting Matter" under
contract 2004 No R113-CT-2004-506078.

\end{document}